\documentclass[fleqn,twoside]{article}
\usepackage{espcrc2}

\usepackage{graphicx}
\usepackage{epsfig}

\newcommand{\AmS}{{\protect\the\textfont2
  A\kern-.1667em\lower.5ex\hbox{M}\kern-.125emS}}

\newcommand{\la}{\langle}
\newcommand{\ra}{\rangle}
\newcommand{\cC}{{\cal C}}

\newcommand{\Z}{{Z \!\!\! Z}}
\newcommand{\be}{\begin{equation}}
\newcommand{\ee}{\end{equation}}
\newcommand{\beqn}{\begin{eqnarray}}
\newcommand{\eeqn}{\end{eqnarray}}

\newcommand{\next}{n_{\mathrm{ext}}}
\newcommand{\nint}{n_{\mathrm{int}}}
\title{
\thispagestyle{empty}
\vspace{-35mm}
\rightline{\small LU-ITP 2001/22~~~~~}
\rightline{\small RNCP-Th01025~~~~~}
\rightline{\small KANAZAWA-01-11~~~~~}
\rightline{\small 10 October, 2001~~~~~}
\vspace{10mm}
      Monopoles, confinement and deconfinement in lattice compact QED
      in (2+1)D with external fields
}

\author{M.~N.~Chernodub\address{ITEP, B.Cheremushkinskaya 25, Moscow, 117259,
        Russia and \newline
        Institute for Theoretical Physics, Kanazawa University, 
        Kanazawa 920-1192, Japan}
        \thanks{M. N. Ch. is supported by JSPS Fellowship P01023.},
        E.-M.~Ilgenfritz\address[osa]{Research Center for Nuclear Physics,
        Osaka University, Osaka 567-0047, Japan}%
        \thanks{E.-M. I. thanks for the support by the Ministry
        of Education, Culture and Science of Japan (Monbu-Kagaku-sho)
        and for a CERN visitor grant.}
        A.~Schiller\address{Institut f\"ur Theoretische Physik,
        Universit\"at  Leipzig, D-04109 Leipzig, Germany}
        \thanks{Presented by A. S. at Lattice'01.}
}
       
\begin{document}

\begin{abstract}
Finite temperature compact electrodynamics in (2+1) dimensions is
studied in the presence of external electromagnetic fields. 
The deconfinement temperature is found to be insensitive 
to the external fields. This result corroborates our
observation that external fields create additional small--size magnetic 
dipoles from the vacuum which do not spoil the confining properties 
of the model at low temperature.  
However, the Polyakov loop is 
not an 
order parameter of confinement.
It can vanish in deconfinement in the presence of external field.
This does not mean the restoration of confinement for
certain external field fluxes. As a next step in the study of (2+1)D QED,
the influence of monopoles on the photon propagator is studied.
First results are presented showing this connection in the confining
phase (without external field).
\end{abstract}

\maketitle

\section{INTRODUCTION AND MODEL}
In $3D$ compact QED confinement is proven and understood.
The confinement is due to presence of Abelian monopoles --
topological defects which appear due to the compactness of the gauge
group. The confining property is lost at sufficiently high temperature.
In~\cite{Chernodub:2001ws} we have demonstrated that the monopoles are
sensitive to the confinement-deconfinement transition.
In the confinement phase the monopoles are in the plasma state while 
in the deconfinement phase the monopoles appear in the form of a dilute 
gas of magnetic dipoles. Here we present results on the influence of
an external electromagnetic field on the confining and monopole properties of
the compact Abelian gauge model in 
(2+1)D~\cite{Chernodub:2001da}.

The action with an external field is given by~\cite{DamgaardHeller}
\be
S[\theta, \theta^{\mathrm{ext}}] = - \beta \sum\nolimits_p 
\cos\Bigl(\theta_p - \theta^{\mathrm{ext}}_p\Bigr)
\label{eq:extint}
\ee
with
$\beta = 1/(a\, g^2_3)$,  $T/g^2_3 =\beta/L_t$.
The non-zero quantized external electric ($E$) or magnetic field ($B$) is directly
coupled to plaquettes in the $31$ or $12$ plane, respectively ($n_{E/M} \in \Z$):
\beqn
 E & = & \theta^{\mathrm{ext}}_{31}= {2\pi n_{E}}/ (L_{3}\, L_{1})\, \nonumber \\
 B & = & \theta^{\mathrm{ext}}_{12}= {2\pi n_{M}
}/ (L_{2}\, L_{1})\,.
  \label{theta:quantE}
\eeqn
Due to the form of action the maximal number of external flux quanta is restricted.
We used 
a Monte Carlo algorithm, which combines a local Monte Carlo step (Metropolis and 
microcanonical sweep) with a global update step changing the internal field 
by a number of flux units, in order to improve ergodicity.
\vspace{-2mm}

\section{POLYAKOV LOOP AND STRING TENSION}
\vspace{-2mm}
Usually, the Polyakov loop $L(\mathbf x)$ is used to probe confinement.
Due to its Abelian nature  
the correlator of two Polyakov loops in the presence of a external field 
$\vec F^{\mathrm{ext}}$ (we use $\vec F^{\mathrm{ext}}=(0,E,0)$ or $(0,0,B)$)
can be written as: 
\be
{\la L(\mathbf 0) L^{*}(\mathbf R)\ra }_{\vec F^{\mathrm{ext}}} 
\propto
{\mathrm{e}}^{ i
     \Phi_\cC(\vec F^{\mathrm{int}})
- L_t \, V(\mathbf R; {\vec F^{\mathrm{ext}}} )
  }  \, ,
\label{ll:H}
\ee
where $V$ is the potential and $\Phi_\cC$ 
is the non-vanishing flux of the internal field $F$ which 
penetrates the surface spanned on the contour $\cC$ given by the test 
particle 
trajectories.

Also the internal
fluxes through the corresponding planes are quantized.
Only the internal electric field 
contributes to the flux $\Phi_\cC(\vec F^{\mathrm{int}})$.

\begin{figure}[!htb]
\vspace{-10mm}
    \epsfxsize=6.6cm \epsffile{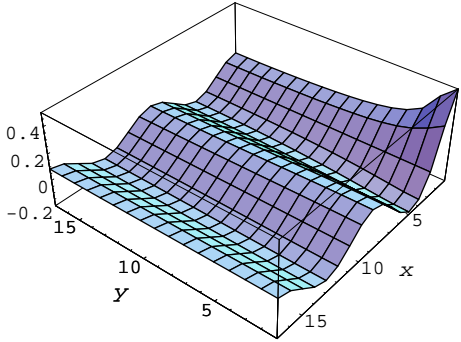}
\vspace{-12mm}
\caption
{Real part of the Polyakov loop correlator in the $x$-$y$
  plane for
  $n_E=4$ in deconfinement phase ($\beta=2.6$). Half of a lattice $32^2
\times 8$ is   shown.}
\label{fig:1}
\end{figure}

An external {\it magnetic} field is directed along the Polyakov loop,
therefore, the induced internal field does not contribute to $\Phi_\cC$.
Consider the correlator in an external {\it electric} field
\be
  {\la L(0,0) \, L^{*}(x,y)\ra }_E
\propto
   {\mathrm{e}}^{2 \pi i x \nint/L_s   - L_t \,
  V(x,y;E)}
\label{ll:H:lat}
\ee
with $\Phi_\cC = 2 \pi \, \nint \, x \, \slash L_s$.
Our numerical results suggest that the internal fluxes  can adequately be
described by taking into account only the    
{\it most probable} flux state $\nint=\nint(\next,\beta,L_i)$.
The oscillating part of the correlator is defined by the electric component
of the internal field.
The correlator (Fig.~\ref{fig:1})
\vspace{-6mm}
simultaneously characterizes both the screening of the external field
(phase factor) and the potential of the test electric charges (by its modulus).

To evaluate the string tension $\sigma$, we use two Polyakov ``plane''
operators which are defined as a sum of Polyakov loops along and
perpendicular to the external field. It can be shown that in the presence of
an electric field
\beqn
&&  {\la L_\parallel(0) L^*_\parallel(x)\ra}_{n_E}  =
  \label{ll:parallel:2} \\
&&
{\mathrm{const}} \cdot
  {\mathrm e}^{2 \pi i n_{\mathrm{int}}\, x \slash L_s } \, \cosh
  \Bigl[{\sigma}\, L_t \, \Bigl(x - \frac{L_s}{2}\Bigr)\Bigr] \,,
\nonumber
\eeqn
therefore, the plane--plane
Polyakov loop correlator parallel to the electric field oscillates with
a decreasing amplitude.

\begin{figure}[!htb]
\vspace{-5mm}
\epsfxsize=6.0cm\epsffile{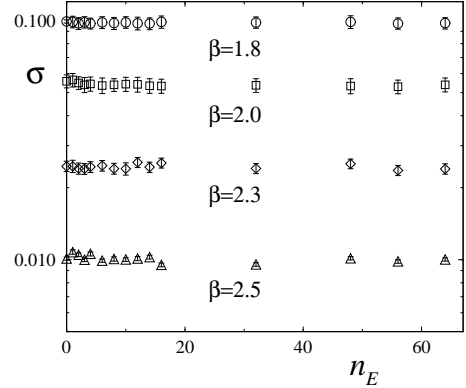}
\vspace{-9mm}
\caption{Fitted string tension for various $\beta$ values as
function $n_E$ (r).}
\label{fig:2}
\vspace{-6mm}
\end{figure}

The plane--plane correlator perpendicular to the field decreases
exponentially (without oscillations) as function of the distance between
the planes:
\beqn
 && {\la L_\perp(0) L^*_\perp(y)\ra}_{n_E}  =
  \label{ll:perp:2}\\
&& 
{\mathrm{const}} \cdot
  \cosh \Bigl[{\sigma_{\mathrm{eff}}}(\sigma,\nint)\, L_t \,
  \Bigl(y - \frac{L_s}{2}\Bigr)\Bigr]\,,
\nonumber
\eeqn
\be
  \sigma_{\mathrm{eff}}=
  \frac{1}{L_t} \,
  {\mathrm{arccosh}} \Bigl[
  \cosh(\sigma\, L_t) - \cos \frac{2 \pi \nint}{L_s} + 1 \Bigr] .
  \label{eff:sigma}
\ee
The essential 
lesson is that the effective string tension ${\sigma_{\mathrm{eff}}}$ does not tell
anything about  confinement properties described by ${\sigma}$.
The fitted string tension (Fig.~\ref{fig:2}), 
the confinement property and the phase structure are not changed 
due to the presence of the external fields.

\begin{figure}[!htb]
\epsfxsize=6.5cm \epsffile{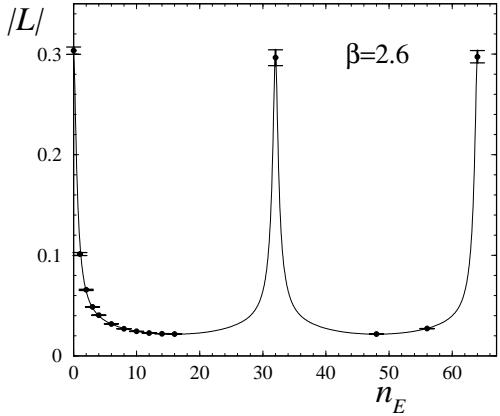}
\vspace{-8mm}
\caption{ The absolute value of the bulk Polyakov loop
  {\it vs.} the external electric flux $n_E$ and its fit
 in the deconfinement phase.}
\label{fig:3}
\vspace{-7mm}
\end{figure}

On the other side, in the case of a non-vanishing $E$, the Polyakov
loop expectation value may vanish due to its tree level contribution
regardless of the value of the actual string tension (see Fig.~\ref{fig:3}).

\vspace{-3mm}
\section{MONOPOLE PROPERTIES AND THE PHOTON PROPAGATOR}
\vspace{-2mm}
Now we present a few results based on a cluster analysis
of the monopole configurations~\cite{Chernodub:2001ws}. 
We observe that the plasma component of the single monopole ensemble
does not feel the external field. 
This is in agreement with the observation made before that the
confinement property does not
depend on the external field.

On the contrary, the dipole density changes drastically as the external field
increases (Fig.~\ref{fig:4}):
\begin{figure}[!htb]
\vspace{-2mm}
    \epsfxsize=7.0cm\epsffile{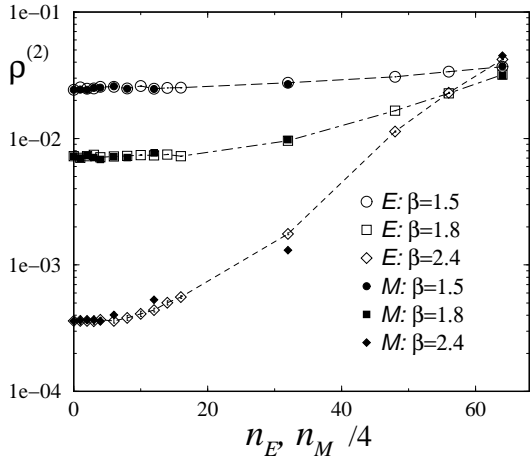}
\vspace{-8mm}
\caption{Clusters of two
monopoles (dipoles) {\it vs.} $n_{E/M}$ (r) for some $\beta$ values.}
\label{fig:4}
\vspace{-6mm}
\end{figure}
the field creates additional monopoles  
popping up as dipoles from the vacuum. The larger the temperature (or $\beta$),
the larger the increase of the dipole density.
With a non--zero electric field, the system is anisotropic in all directions.
As the external field increases, the dipoles become elongated,
increasingly with the external field, along the direction of the applied field
while in an external magnetic field dipoles become more polarized along the
(temporal) $z$--direction similar to increasing temperature.

\begin{figure}[!htb]
\vspace{-5mm}
\epsfxsize=6.5cm\epsfysize=6.0cm\epsffile{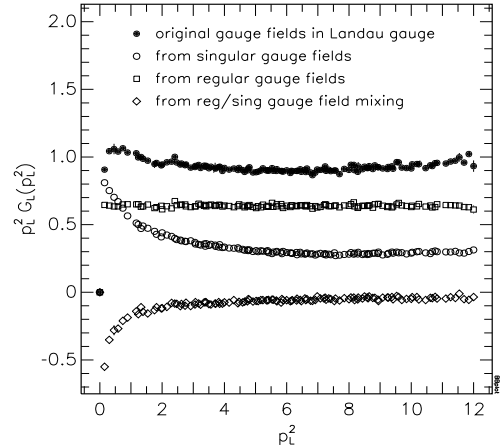}
\vspace{-10mm}
\caption{Different contributions to
$T=0$ photon propagator $G(p_L^2)$ in Landau gauge {\it vs.} lattice
momentum squared, $p_L^2$ (lattice $16^3$, $\beta=2.0$).}
\label{fig:5}
\vspace{-6mm}
\end{figure}

At zero temperature the model is confining.  It is interesting how this is
reflected in the photon propagator and how monopoles are showing up there.  
In Fig.~\ref{fig:5} we show in lattice momentum space how singular
(monopole-like) and regular (photon-like) lattice gauge field modes
contribute to the photon propagator at zero temperature. The propagator has
been measured in Landau gauge taking into account Gribov copies and removing
zero momentum modes. Details 
will be presented elsewhere~\cite{CIS2001c}. 
Similar 
investigation of the influence of center vortices on the gluon propagator 
was done in Ref.~\cite{Langfeld}.

\end{document}